\documentclass[twocolumn,showpacs,preprintnumbers,amsmath,amssymb,prl]{revtex4}

\usepackage{graphicx}
\usepackage{dcolumn}
\usepackage{bm}

\begin{document}

\title{The transverse breathing mode of an
elongated Bose-Einstein condensate}
\author{F. Chevy, V. Bretin, P. Rosenbusch, K. W. Madison, and J. Dalibard}
\affiliation{Laboratoire Kastler Brossel$^*$, 24 rue Lhomond,
75005 Paris, France}
\date{Received November 23, 2001}

\begin{abstract}

{We study experimentally the transverse monopole mode of an
elongated rubidium condensate. Due to the scaling invariance of
the non-linear Schr\"odinger (Gross-Pitaevski) equation, the
oscillation is monochromatic and sinusoidal at short times, even
under strong excitation. For ultra-low temperatures, the quality
factor $Q=\omega_0/\gamma_0$ can exceed 2000, where $\omega_0$ and
$\gamma_0$ are the mode angular frequency and damping rate. This
value is much larger than any previously reported for other
eigenmodes of a condensate. We also present the temperature
variation of $\omega_0$ and $\gamma_0$.}

\end{abstract}


\maketitle

Collective excitations constitute one of the main sources of
information for understanding the physics of many-body systems.
Since the discovery of Bose-Einstein condensation (BEC) in weakly
interacting bosonic gases \cite{Varenna98} much theoretical and
experimental effort has been devoted to the study of the low
energy excitations of these systems
\cite{Stringari96,Jin96,Mewes96,Onofrio00,Marago00}. At ultra-low
temperatures, the measured frequencies of these excitations are
found to be in very good agreement with the ones derived from a
simple linearization of the Gross-Pitaevskii equation, which
describes the evolution of the condensate wave function.

The temperature dependence of the damping and frequency of these
oscillations has attracted strong attention of both experimental
\cite{Jin97,Stamper98,Marago01} and  theoretical groups
\cite{Dalfovo99,Guilleumas99,Rusch99,AlKhawaja00,Giorgini00,Reidl00,Morgan00,Kagan01,Jackson01,Williams01,Jackson01b}.
Indeed, a dilute BEC is a very convenient system for the study of
dissipation in a macroscopic quantum object. At present, there
still exist significant discrepancies between some available
experimental results and the corresponding theoretical predictions
(see e.g. \cite{Reidl00,Jackson01b}). Therefore new measurements
on simple and well identified modes are clearly needed to test the
various theoretical approaches.

It has been known for some time that in an isotropic trap, the
monopole (or breathing) oscillation mode may exhibit very unusual
features. For instance, Boltzmann noticed that, within the
framework of the equation which bears his name, the breathing mode
of a set of particles confined in a 3D isotropic harmonic trap is
undamped, irrespective of the ratio between the average collision
rate and the trap frequency \cite{Cercignani,Guery99a,Ghosh00}.
More recently it was pointed out that in a 2D isotropic harmonic
trap, the breathing mode of identical bosons interacting via a
contact potential is also undamped \cite{Kagan96,Pitaevskii97}.
This property, which holds at any temperature, can be related to
the existence of a hidden symmetry of the problem described by the
two-dimension Lorentz group SO(2,1) \cite{Pitaevskii97}.

The purpose of the present work is to investigate experimentally
the transverse breathing mode of an elongated condensate. We find
that this mode exhibits very unique features, such as an extremely
large quality factor $Q=\omega_0/\gamma_0\sim 2000$ for
temperatures much lower than the condensation temperature. Here
$\omega_0$ and $\gamma_0$ are the angular frequency and damping
rate, respectively. This quality factor, much larger than any
previously reported value for an oscillation mode of a gaseous
BEC, should provide a sensitive test of the various theoretical
models of dissipation at ultra-low temperature. Another very
unusual property lies in the fact that the measured frequency
$\omega_0$ is nearly independent of temperature. Furthermore, we
find that the transverse area of the condensate undergoes a pure,
isochronal oscillation even for strong excitation. This striking
behavior emerges from a scaling invariance of the time-dependent
Gross-Pitaevski equation.

Our $^{87}$Rb condensate is formed by radio-frequency (rf)
evaporation of $10^9$ atoms in a Ioffe-Pritchard magnetic trap.
The initial temperature of the cloud pre-cooled using optical
molasses is 100 $\mu$K. We vary the final temperature of the gas
and the number of atoms in the condensate by adjusting the final
rf, $\nu_{\rm f}$.  The condensation threshold is reached at
$T_c=290$~nK, with $N_c=8.4\;10^5$~atoms \cite{calibration}. The
coldest sample reliably reproducible ($T\sim 40$~nK) is a
quasi-pure condensate with $N_0=10^5$~atoms. It is obtained using
$\nu_{\rm f}=\nu_0 +6$~kHz, where $\nu_0$ is the rf at which the
trap is emptied.

The trap has a longitudinal frequency
$\omega_z/(2\pi)=11.8\;(1)$~Hz for the atoms prepared in the
stretched state $m=2$ of the $5\,S_{1/2},\; F=2$ ground state. The
transverse magnetic gradient is $b_\bot'=1.33$~T\,m$^{-1}$ and the
transverse frequency $\omega_\bot$ is adjusted by varying the bias
field $B_0$ at the center of the trap. We choose $B_0 =
0.844\;10^{-4}$~T, which leads to $\omega_\bot/(2\pi)
=182.6\;(4)$~Hz \cite{anisotropy}. Both frequencies $\omega_z$ and
$\omega_\bot$ are measured by monitoring the oscillation of the
center-of-mass of the atom cloud.

Once the condensate is formed, we excite the transverse breathing
mode by modifying the bias field from $B_0$ to $B'_0$, which
changes the frequency $\omega_\bot$ to $\omega_\bot'>\omega_\bot$.
After a time $\tau \ll 2\pi/\omega_\bot'$, we set the bias field
back to its original value. We let the cloud oscillate freely in
the magnetic trap for an adjustable time $t$ and then measure the
transverse density profile of the condensate after a period of
free expansion. In this pursuit, we suddenly switch off the
magnetic field, allow for a 25~ms free-fall, and image the
absorption of a resonant laser by the expanded cloud. The imaging
beam propagates along the $z$ axis. We fit the density profile of
the condensate assuming the parabolic shape of the Thomas-Fermi
distribution and extract the radii $R_x$ and $R_y$ in the plane
$z=0$. The temperature is obtained from a Gaussian fit of the
uncondensed part of the distribution.

Within the Thomas-Fermi approximation, the evolution of the
trapped condensate after the excitation of the transverse
breathing mode is well described by a time-dependent scaling
transformation \cite{Kagan96,Castin96,Kagan97}.  The spatial
density $\rho({\bf r},t)$ at time $t$ is deduced from the spatial
density at time $0$ (i.e. just after the transverse frequency is
set back to its value $\omega_\bot$) by:
\[
\rho(x,y,z,t)=\frac{1}{\lambda_\bot^2 \lambda_z}\;
\rho\left(\frac{x}{\lambda_\bot},\frac{y}{\lambda_\bot},
\frac{z}{\lambda_z},0 \right)\ .
\]
The evolution of the scaling parameters $\lambda_\bot$ and
$\lambda_z$ is given by \cite{Castin96,Kagan97}:
\begin{equation}
\ddot \lambda_\bot=\frac{\omega_\bot^2}{\lambda_\bot^3\lambda_z}
-\omega_\bot^2 \lambda_\bot \qquad \qquad \ddot
\lambda_z=\frac{\omega_z^2}{\lambda_\bot^2\lambda_z^2} -\omega_z^2
\lambda_z \ . \label{scaling}
\end{equation}

\begin{figure}
 \hskip -0.0cm
 \centerline{\includegraphics{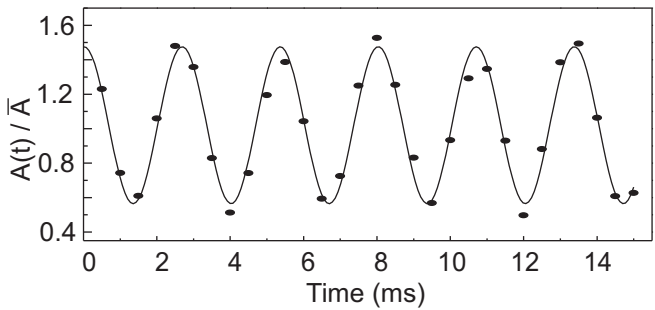}}
 \vskip -0.2cm
 \caption{Variation of $A(t)=R_x^2(t) +R_y^2(t)$,
 after a strong excitation of the transverse
breathing mode. The line is a sinusoidal fit of the data
($\omega'_\bot/2\pi=250$~Hz, $\tau=800\;\mu$s).}
 \label{fig:strongex}
 \vskip -0.5cm
\end{figure}

We start our analysis with the evolution for relatively short
times $t\ll 2\pi/\omega_z$. In this case one can neglect the
variation of $\lambda_z$ and integrate the equation for
$\lambda_\bot$:
\[
\lambda_\bot(t)=\left(
\cosh(\alpha)+\sinh(\alpha)\sin(2\omega_\bot t-\beta)
\right)^{1/2}\ ,
\]
where the parameters $\alpha$ and $\beta$ depend on the initial
conditions. One remarkable consequence of this solution is that
the quantity $A(t)=R_x^2(t) +R_y^2(t) = \lambda_\bot^2(t)\,A(0)$
always undergoes a sinusoidal oscillation at frequency
$2\omega_\bot$, irrespective of the strength of the excitation
\cite{TOF}. This is confirmed by the results of fig.
\ref{fig:strongex}, where we observe a sinusoidal oscillation of
$A(t)$ with a factor 3 between the maxima and the minima. The
frequency of the oscillation is indeed found to be equal to
$2\,\omega_\bot$, within our experimental uncertainty ($\pm
0.5$~\%).

At longer times one has, in principle, to take into account the
variation of $\lambda_z$. We restrict our analysis to small
amplitude oscillations and write $\lambda_i=1+\epsilon_i$, where
$\epsilon_\bot,\epsilon_z \ll 1$. The linearized equations of
motion
\[
\ddot \epsilon_\bot + \omega_\bot^2 \left(4\epsilon_\bot
+\epsilon_z \right)=0 \qquad  \ddot \epsilon_z + \omega_z^2
\left(2\epsilon_\bot +3\epsilon_z \right)=0
\]
give rise to two modes \cite{Stringari96,Mewes96}. One is a fast
mode of frequency $\sim 2\omega_\bot$ corresponding to
$\epsilon_\bot/\epsilon_z=\omega_\bot/\omega_z$. The other one is
a slow mode of frequency $\sqrt{5/2}\;\omega_z$ with
$\epsilon_\bot/\epsilon_z=-1/4$. Our excitation scheme corresponds
to $\dot \epsilon_\perp\neq 0$ at time 0 while the three other
quantities $\epsilon_\bot$, $\epsilon_z$ and $\dot \epsilon_z$ are
approximately zero at initial time. Under these conditions, we
mainly excite the fast mode $2\omega_\bot$ since the relative
weight of the slow mode in the evolution of $\lambda_\bot$ is only
$\omega_z/(\sqrt{40}\, \omega_\bot)<0.01$. For this fast mode, the
condensate dynamics consists essentially in a transverse monopole
oscillation since $\epsilon_z \ll \epsilon_\bot$.

\begin{figure*}
 \hskip -0.0cm
 \scalebox{0.995}{\centerline{\includegraphics{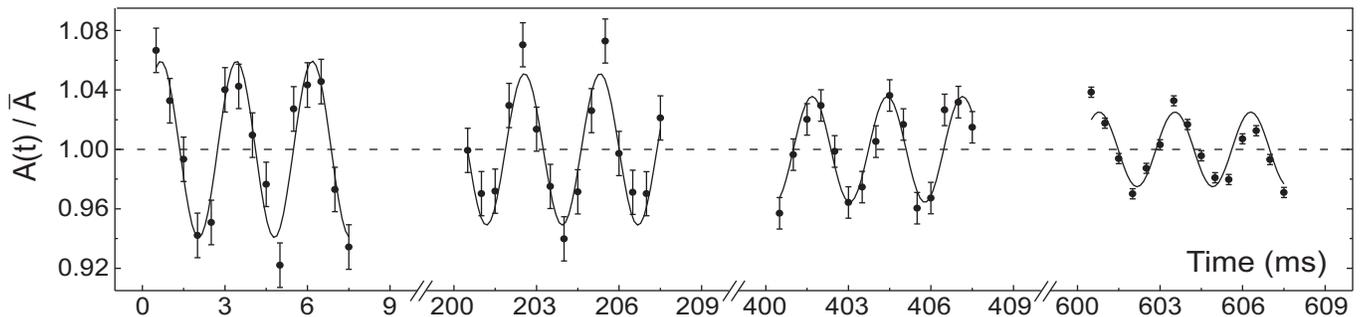}}}
 \vskip -0.2cm
 \caption{ Long time evolution of $A(t)/\bar A$ at low
temperature ($40$~nK), where $\bar A$ is the average of $A(t)$
over each time interval [0,9~ms], [200~ms,209~ms] ... The
oscillation is well fitted by a damped sinusoid. Here
$\omega'_\bot/2\pi=230$~Hz and $\tau=75\;\mu$s. }
 \label{fig:weakex}
 \vskip -0.5cm
\end{figure*}

We investigate the oscillation and the damping of this mode on a
long time scale (up to $0.7\;$s). An example is given in fig.
\ref{fig:weakex} for $T=40\; (\pm 20)$~nK (note that for such cold
clouds, corresponding to a quasi-pure condensate, the temperature
can only be inferred from the final rf used for the evaporation).
The evolution of $A(t)$ is well fitted by a damped sinusoidal
function, $A_0+\delta A_0\,\cos(\omega_0 t+\phi)\;e^{-\gamma_0
t}$, with $\delta A_0/A_0=0.063\;(4)$,
$\omega_0\,/\,2\pi=366.3\;(5)$~Hz and $
\gamma_0=1.2\;(2)$~s$^{-1}$. This corresponds to a quality factor
$Q=\omega_0/\gamma_0 \sim 2000$ much larger than any previously
reported for other eigenmodes of a BEC. For instance, measurements
performed with a TOP trap ($\omega_z=2\sqrt{2}\;\omega_\bot$) gave
 $Q \sim 200$ for the lowest $m=0$ mode \cite{Jin97}. For a
cigar shape trap, measurements of the low frequency mode discussed
above ($\omega=\sqrt{5/2}\;\omega_z$) led to $Q\simeq 80$
\cite{Stamper98}. A smaller value ($Q\sim 25$) was measured for
the scissors mode \cite{Marago01}. The present quality factor is
also one order of magnitude larger than what we find for the
transverse quadrupole mode under the same experimental conditions
\cite{Chevy00}.

\begin{figure}
 \hskip -0.1cm
 \centerline{\includegraphics{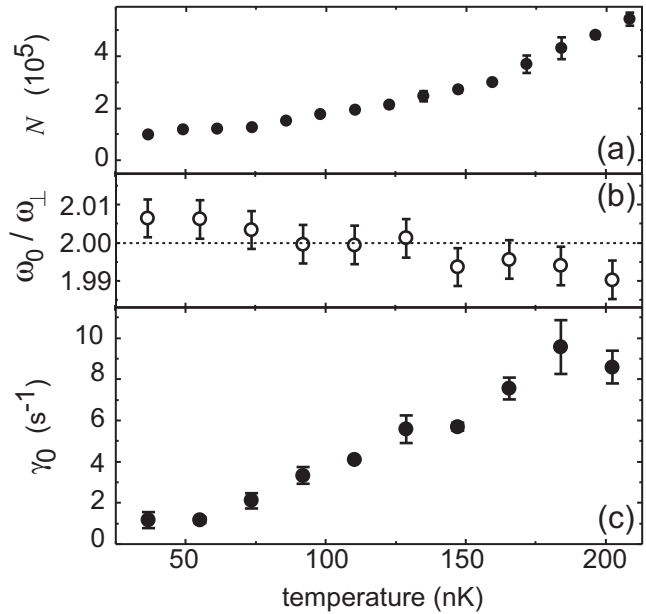}}
 \vskip -0.2cm
 \caption{(a) Total number of atoms $N$,
(b) breathing mode frequency $\omega_0/\omega_\bot$ and (c)
damping rate $\gamma_0$, as a function of temperature. Each point
in (b) and (c) corresponds to $\nu_{\rm f}-\nu_0=$ 6~kHz, 9~kHz,
$\ldots$, 33~kHz ($\omega'_\bot/2\pi=230$~Hz, $\tau=150\;\mu$s).}
 \label{fig:variationwithT}
 \vskip -0.5cm
\end{figure}

We measure the oscillation frequency and the damping rate over a
large temperature range compatible with the detection of a
condensate. The results are given in
fig.~\ref{fig:variationwithT}, together with the total number of
atoms $N$. The measurement of $\gamma_0$ is made from the decay of
the oscillation amplitude between $0$ and $200$~ms. For the points
of lowest and highest temperature in figs. 3b and 3c, we
calculate, including mean field and finite size corrections
\cite{Dalfovo99}, $T/T_c=0.3$ and $0.8$, respectively.

The frequency $\omega_0$ is found to be very close to the
Thomas-Fermi prediction $2\omega_\bot$, and it varies only
slightly with temperature (fig. 3b). In particular, we do not
observe the 10\;\% relative decrease predicted in
\cite{Giorgini00} when $T/T_c$ increases from 0 to 0.8.

The damping rate $\gamma_0$ varies approximately linearly with
temperature in the range $50-200$~nK.  The linear variation of
$\gamma_0$ for $k_{\rm B}T$ larger than $\hbar \omega_0$ ($\sim
20$~nK) and than the chemical potential $\mu$ ($\sim 60$~nK for
the coldest point in fig.~\ref{fig:variationwithT}) is expected if
one assumes either Beliaev-type or Landau-type damping of the
monopole excitation (see e.g. \cite{Fedichev98}). In the Beliaev
mechanism, the monopole excitation decays into two excitations
with a lower energy. For Landau damping, it annihilates in a
collision with a thermal excitation to give rise to another
excitation. The calculation of the corresponding damping rate is
quite complex for inhomogeneous systems and can lead to
drastically different results for modes with similar energies,
depending on the stochastic or regular nature of the relevant
thermal excitations \cite{Fedichev98}. In this respect it is worth
noting that the frequency of this transverse breathing mode is an
integer multiple of the trap oscillation frequency, which is
usually not the case for other modes of the condensate.

\begin{figure}
 \hskip -0.35cm
 \centerline{\includegraphics{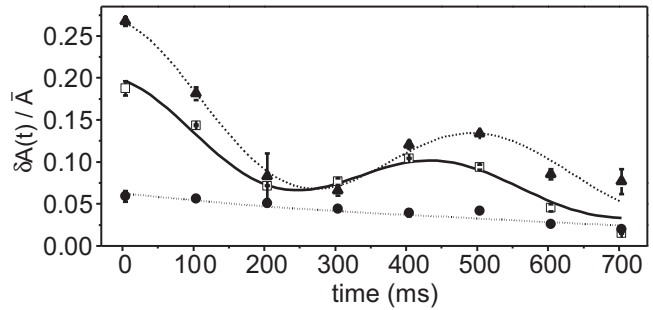}}
 \vskip -0.2cm
\caption{Long term evolution of the amplitude of the breathing
mode for different excitation strengths
($\bullet$:$\tau=75\,\mu$s, $\square$: $\tau_1=300\,\mu$s,
$\blacktriangle$:$\tau_2=450\,\mu$s). Each point at $t=0$, 100~ms,
$\ldots$, 700~ms represents the amplitude of the breathing mode
measured during the period $\left[t,t+8~{\rm ms}\right]$. The fit
function is indicated in the text. Here
$\omega'_\bot/2\pi=230$~Hz.}
 \label{fig:nonexponential}
 \vskip -0.5cm
\end{figure}

For a very low temperature ($T\sim 40$~nK) and an increasing
excitation strength, we observe a significant deviation of the
long term behavior of $A(t)$ with respect to the simple damped
sinusoid shown above. Fig. \ref{fig:nonexponential} shows the
oscillation amplitude for the weak excitation ($\bullet$,
$\tau=75\;\mu$s) together with two sets of data obtained for
$\tau_1=300\;\mu$s ($\square$) and $\tau_2=450\;\mu$s
($\blacktriangle$). One clearly sees that for strong excitation
the amplitude of the oscillations passes through a local minimum
around $t=300$~ms before increasing again, and decaying finally
after 700~ms. We fit our experimental data with the
phenomenological function $A(t)=A_0 + \delta A(t)\, \cos(\omega_0
t+\phi)$ where the amplitude $\delta A(t)$ is now slowly modulated
in time, in addition to the exponential decay already discussed:
$\delta A(t)=\delta A_0\, \left(1 -\xi \sin^2 (\delta \omega\, t)
\right) \;e^{-\gamma t} $. The best fit parameters are for these
two cases $\gamma_{1}=1.48\; (15)$~s$^{-1}$, $\delta
\omega_{1}/(2\pi)=1.08\;(4)$~Hz, $\xi_1=0.5\,(1)$, and
$\gamma_{2}=1.35\; (15)$~s$^{-1}$, $\delta
\omega_{2}/(2\pi)=0.96\;(4)$~Hz, $\xi_2=0.6\,(1)$.

Two scenarios can be invoked to explain this evolution of $A(t)$,
which is not predicted by (\ref{scaling}). A first explanation
could consist in assuming that we actually excite two modes whose
frequencies $\sim \omega_0$ differ by a small amount and that we
observe the beating between these modes. However, if we assume
that these two modes are excited linearly, the relative amplitude
of the beating should not depend on the excitation strength,
contrary to the observed. Furthermore, this explanation is not
confirmed by a linear analysis of the time dependent
Gross--Pitaevskii equation for our experimental conditions
\cite{CastinPrikopenko}. A second, more plausible, scenario
involves a non-linear process, in which the monopole excitation
may be converted in a reversible way into a sum of other
excitations of frequency $\omega_j$. This process is efficient if
it is resonant: $\omega_0 \simeq \sum_j \omega_j$. In this
context, the modulation in fig. \ref{fig:nonexponential} can be
interpreted as a parametric oscillation.

As a possible candidate for such a parametric (or Beliaev type)
oscillation \cite{Hodby01}, one may consider the excitation of the
pair of modes $l=2,m=\pm 1$ \cite{Stringari96,DGO99,Marago00}. The
frequency of each mode is $(\omega_\perp^2+\omega_z^2)^{1/2}$ so
that the resonance condition given above is indeed nearly
fulfilled. By imaging the condensate along the $y$ direction, we
observe a weak excitation of this mode. However, no increase in
its amplitude is found around $t=$300~ms when the amplitude of the
breathing mode collapses (cf. fig. \ref{fig:nonexponential}).

Alternatively one may consider the excitations consisting in a
pair of phonon-like excitations propagating along $z$ with
momentum $k$ and $-k$, each with an energy $\hbar \omega_0/2\sim
\hbar \omega_\bot$ \cite{Kagan01}. To test this latter hypothesis,
we perform the same experiment for various values of
$\omega_\perp$, so that it spans the whole interval between two
successive phonon modes (typically 6~Hz in this energy domain).
However, expecting a resonance when $\omega_0/2$ coincides with
the frequency of a phonon mode, we do not observe any significant
change in the evolution of $A(t)$.

In any case we note that for the long oscillation times involved
here, small non-linear terms in the confining potential (varying
as $x^2 z$ and $y^2 z$) may play a significant role. Further work
investigating this possibility is under progress in our
laboratory.

In summary we have presented in this letter a study of the
transverse breathing mode of an elongated Bose-Einstein
condensate. For low temperatures and short evolution time, the
observed isochronal and sinusoidal oscillation is in excellent
agreement with the prediction based upon the scaling invariance of
the Gross-Pitaevski equation. However, the temperature dependence
of the frequency and damping of the oscillation reveals, on the
contrary, some unexpected features. First, the observed damping is
comparatively ten times smaller than what was previously reported
for other modes of a BEC (quality factor of 2000 instead of
100-200). Second, the frequency of the breathing mode is found to
be quasi-independent of temperature, in contradiction with recent
theoretical predictions. Finally the long time evolution of this
mode when strongly excited reveals an unexpected non-linear mixing
phenomenon. This could be explored further using the formalism of
\cite{Castin97}, by looking at a possible dynamical instability of
the Gross-Pitaevskii equation in this regime of strong excitation.

{\acknowledgments We thank Y. Castin, Yu. Kagan, L. Pitaevski, and
G. Shlyapnikov for useful discussions. K. M. and P. R. acknowledge
support by DEPHY and Alexander von Humboldt-Stiftung,
respectively. This work was partially supported by CNRS,
Coll\`{e}ge de France and DRED.}


\begin{thebibliography}{99}
\bibitem{calibration}
The value of $T_c$ is measured by the time-of-flight method. We
calibrate $N_c$ using the prediction relating $N_c$ and $T_c$,
including mean field corrections \cite{Dalfovo99}. The result for
$N_c$ is in excellent agreement with a direct determination based
upon the optical thickness of the cloud.
\bibitem{anisotropy}
More precisely, we measure $\omega_x/(2\pi)=183.2 (3)$~Hz and
$\omega_y/(2\pi)=182.0 (3)$~Hz. In the following we neglect the
small coupling between the transverse breathing mode (frequency
2$\omega_\bot$) and the quadrupole mode (frequency
$\sqrt{2}\,\omega_\bot$) induced by this slight anistropy.
\bibitem{TOF}
This remains true even when the time-of-flight expansion is taken
into account.
\end{thebibliography}
\end{document}